\documentclass{wlscirep}

\usepackage{amsmath}
\usepackage{siunitx}
\DeclareSIUnit\electrons{e\textsuperscript{$-$}}
\usepackage{changes}

\title{Insights into radiation damage from atomic resolution scanning transmission electron microscopy imaging of mono-layer CuPcCl\textsubscript{16} films on graphene}

\author[1,*]{Andreas Mittelberger}
\author[1]{Christian Kramberger}
\author[1,+]{Jannik C. Meyer}
\affil[1]{Faculty of physics, University of Vienna, Boltzmanngasse 5, 1090 Vienna, Austria}

\affil[*]{andreas.mittelberger@univie.ac.at}
\affil[+]{jannik.meyer@univie.ac.at}


\begin{abstract}
	Atomically resolved images of monolayer organic crystals have only been obtained with scanning probe methods so far. On the one hand, they are usually prepared on surfaces of bulk materials, which are not accessible by (scanning) transmission electron microscopy. On the other hand, the critical electron dose of a monolayer organic crystal is orders of magnitudes lower than the one for bulk crystals, making (scanning) transmission electron microscopy characterization very challenging. In this work we present an atomically resolved study on the dynamics of a monolayer CuPcCl\textsubscript{16} crystal under the electron beam as well as an image of the undamaged molecules obtained by low-dose electron microscopy. The results show the dynamics and the radiation damage mechanisms in the 2D layer of this material, complementing what has been found for bulk crystals in earlier studies. Furthermore, being able to image the undamaged molecular crystal allows the characterization of new composites consisting of 2D materials and organic molecules. 
\end{abstract}

\begin{document}

\flushbottom
\maketitle

\thispagestyle{empty}

\section{Introduction}
Phtalocyanines (Pc), and in particular Copper-Phtalocyanine (CuPc), have attracted great interest in surface science and microscopy, both as an active material~\cite{Xu2015,Noh2017,Yang2012,Mensing2012,Jiang2014} and as a model system. As a model system this class of materials was extensively studied to investigate radiation damage in organic molecules in the transmission electron microscope (TEM)~\cite{Smith1986,Kurata1992,CLARK1979,Clark1980,Fryer1984,Uyeda1972}. It was found that the halogenated versions of these materials were unusually radiation resistant, withstanding doses up to 10 times higher then their non-halogenated counterparts~\cite{Egerton2012}. More recent studies, using aberration corrected (scanning) TEM ((S)TEM), were even able to resolve defects in such molecular crystals~\cite{Haruta2012}. It is, however, important to note that all this was done on crystals with thicknesses of several tens of nanometers~\cite{Haruta2008}. Studies on monolayer organic crystals were so far only possible with scanning probe methods, because they are usually prepared on surfaces that are not accessible by (S)TEM. In this paper we report, to our knowledge, for the first time an atomically resolved image of a monolayer CuPcCl\textsubscript{16} crystal obtained by low-dose imaging combined with translational averaging.

\section{Experimental}
The samples were prepared by evaporating CuPcCl\textsubscript{16} powder purchased from Sigma Aldrich (Phtalocyanine green) onto monolayer graphene on TEM grids (Graphenea). Deposition was carried out in our homemade UHV sample preparation system which allows the transfer of the samples to the STEM in UHV without exposing them to air which is crucial for minimizing contamination. During evaporation, the substrate was held at room temperature and the source was heated to $\SI{400}{\degreeCelsius}$. The nominal film thickness was $\SI{\sim 7}{\angstrom}$ as measured with a quartz microbalance.
All electron microscopy images were acquired in an Nion UltraSTEM 100 at an acceleration voltage of $\SI{60}{\kilo\volt}$ with a beam current of $\SI{\sim 30}{\pico\ampere}$ and a convergence semi-angle of $\SI{33}{\milli\radian}$~\cite{Krivanek2008}. As detector half-angles for the annular dark-field (ADF) detector we used a range of $\SIrange{\sim60}{200}{\milli\radian}$, which is well-suited for imaging of light elements such as carbon~\cite{Krivanek2012}. The electron dose was varied by changing the pixel dwell time and/or the pixel size. Like this, we can achieve doses that are low enough to image the undamaged molecular crystal and are comparable to what is used in biological TEM experiments~\cite{Buban2010}. We acquired images of CuPcCl\textsubscript{16} on graphene with doses ranging from $\SIrange{15}{500000}{\electrons\per\angstrom\squared}$.
Translational averaging was done by cutting the raw image into small subframes, each containing one unit cell of the CuPcCl\textsubscript{16} crystal lattice. Suitable start values for the lattice parameters were found by analyzing the peak positions in the Fourier transform (FT) of the raw image. Those parameters were then manually optimized with visual feedback from the sum of all unit cells of the crystal. For a better visual feedback, the translationally averaged unit cell was periodically repeated which creates a virtual image of the crystal lattice.

\section{Results and Discussion}
\begin{figure}
	\centering
	\includegraphics[width=1.\textwidth]{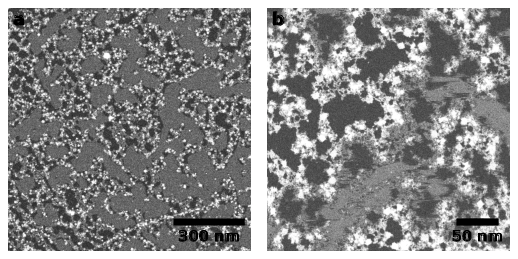}
	\caption{\label{fig:overview}(a) Overview image of a CuPcCl\textsubscript{16} monolayer on graphene. The dark areas are clean graphene, the gray areas the CuPcCl\textsubscript{16} and the bright areas are contamination. The electron dose was $\SI{\sim 15}{\electrons\per\angstrom\squared}$. (b) Higher magnification image of the same sample as in a. The CuPcCl\textsubscript{16} crystal is not stable anymore at this dose ($\SI{\sim 500}{\electrons\per\angstrom\squared}$), which causes the strong dynamics in the gray areas.}
\end{figure}

Figure~\ref{fig:overview}a shows a low-magnification image of the monolayer CuPcCl\textsubscript{16} crystal on graphene. The molecules (gray areas) fill the space in between the contamination (bright areas) very homogeneously. Note that there is no contrast variation visible in the areas covered by the molecules. Also in figure~\ref{fig:overview}b, where the imaging dose was already high enough to damage the molecules and the film is slowly removed, there is no intermediate gray level between a full coverage and the clean graphene. This is a strong indication that we are actually observing a monolayer film of CuPcCl\textsubscript{16} on graphene. It is also supported by earlier publications where the growth of this material was investigated by scanning tunneling microscopy (STM). These works show that the material indeed shows layer-wise growth on graphite surfaces/graphene~\cite{Ludwig1994,Walzer2001,Scheffler2013,Stock2015}. Individual molecules are highly mobile on these surfaces and can only be imaged at very low temperatures (e.g. $\SI{35}{\kelvin}$)~\cite{Dekker1997,Stock2015}. This explains why the molecular crystals are always continuous and there are no individual molecules or small clusters in the center of a clean graphene spot.

\begin{figure}
	\centering
	\includegraphics[width=1.\textwidth]{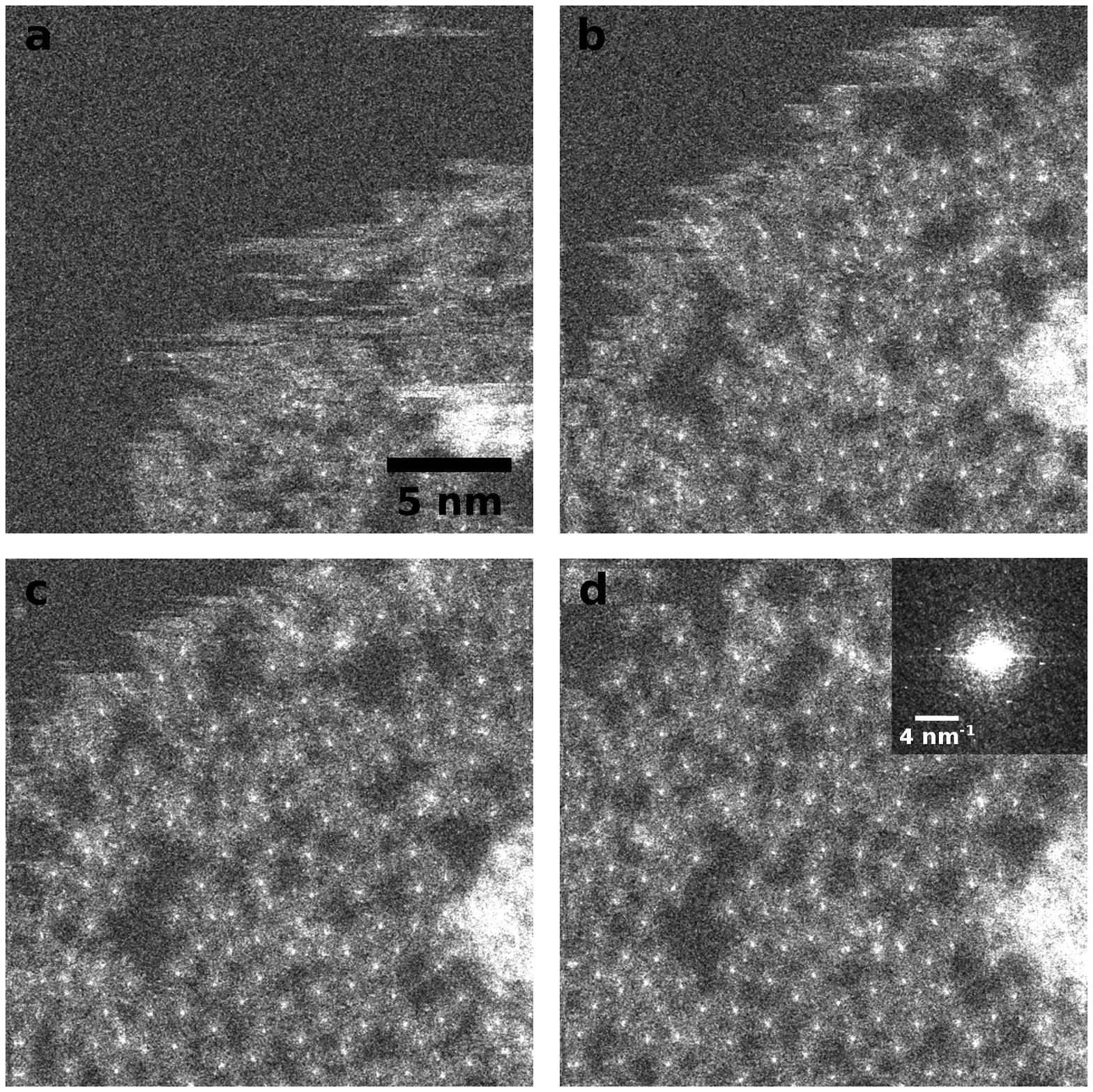}
	\caption{\label{fig:mapping}(a-d) Sequence of 4 images taken in the same spot at a dose of $\SI{\sim 10000}{\electrons\per\angstrom\squared}$. At this dose, the (already damaged) molecules move back into the field of view. Under electron irradiation the structure stabilizes and forms a random 2D-network. Inset in (d): Central part of the power spectrum of the image in (d) that shows the typical reflections of the underlying graphene lattice.}
\end{figure}

Figure~\ref{fig:mapping} shows four consecutive images from the same spot with high resolution and a dose of $\SI{\sim 10000}{\electrons\per\angstrom\squared}$. Here, the molecules move into the previously clean graphene area under electron irradiation. In STEM, horizontal lines in an image indicate movement of the sample, which is clearly visible in panels~a-c. In panel~d, the structure has stabilized and no more movement of the molecules can be observed. Note, however, that at this dose the molecules are already destroyed and what can be seen in the image are only the central copper atoms surrounded by amorphous carbon (which might also include nitrogen atoms from the molecules). This also explains the random network of the molecule leftovers in contrast to the ordered crystal at lower doses, which will be discussed later. The amorphous structure of the sample in this state can also be seen from the power spectrum in the inset of figure~\ref{fig:mapping}d. Apart from the graphene reflections, no crystallinity is visible anymore.
Interestingly, the observed behavior of the molecules on graphene is similar to that of the routinely seen thin and mobile hydrocarbon contamination layer: At intermediate magnifications as in figure~\ref{fig:overview}b the molecules are removed, revealing a perfectly clean graphene surface. At high magnifications as in figure~\ref{fig:mapping}, beam induced deposition and subsequent crosslinking can be observed, yielding a fixed pattern of the deposited material~\cite{Dyck2018}.

These observations provide profound insight into the mechanism of beam-induced deposition of carbon contamination. Such contamination is believed to form from organic molecules on the sample surface that are cracked under the electron beam, and which are constantly replenished by diffusion into the irradiated area along the surface or from the vacuum system~\cite{Ennos1953,Griffiths2010}. In our experiment, we start with a well defined molecule that is present only on the sample surface, and we can identify the fragments of the CuPcCl\textsubscript{16} by the clearly visible central Cu atom that is still present in the deposited amorphous layer. The molecular fragments form a random network, where the separation of the Cu atoms corresponds approximately to the diameter of the molecules. The resulting pattern can be explained as CuPcCl\textsubscript{16} molecules that have lost the chlorine atoms and are interlinked via the thus formed open bonds.  Hence, highly mobile CuPcCl\textsubscript{16} molecules must be present on the graphene surface, besides the localized ones within the crystalline areas. These molecules are destroyed, by scission of the chlorine bond, when they migrate into the beam. Then, the molecule is immobilized if the open bond can connect with existing contamination or other molecules. Indeed, the molecular fragments (as well as the usual carbon contamination, if present) in our experiments always started to grow at existing contamination or defects, never within clean and defect free regions of graphene.
 
\begin{figure}
	\centering
	\includegraphics[width=1.\textwidth]{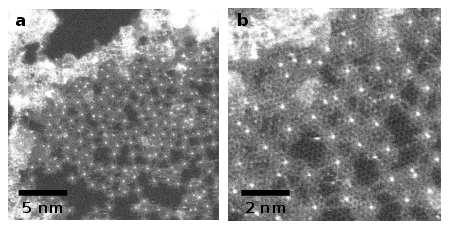}
	\caption{\label{fig:closeup}Even under higher dose irradiation the structure stays stable. The bright dots are the copper atoms from the center of the molecules. They are surrounded by the leftovers of the already degraded molecules which is mostly amorphous carbon. Some of the destroyed molecules are crosslinked by 1D carbon chains (see top-right corner of panel b and figure~\ref{fig:chlorine_loss}a). The doses are $\SI{\sim 30000}{\electrons\per\angstrom\squared}$ and $\SI{\sim 120000}{\electrons\per\angstrom\squared}$ in panels a and b, respectively.}
\end{figure}

\begin{figure}
	\centering
	\includegraphics[width=1.\textwidth]{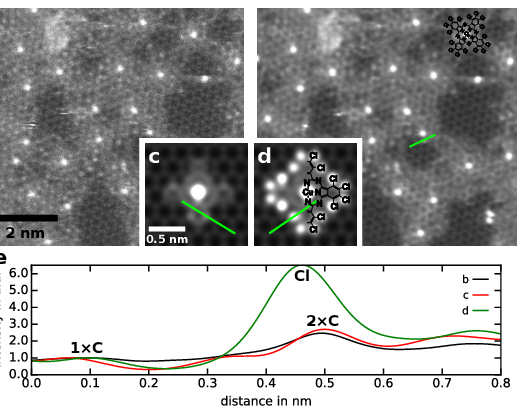}
	\caption{\label{fig:chlorine_loss}(a) Close-up of the structure in figure~\ref{fig:closeup}b showing the cross-linked molecule leftovers. The dose is $\SI{\sim 500000}{\electrons\per\angstrom\squared}$. (b) Same image as in (a) but with shot noise removed by a Gaussian filter with a sigma of $\SI{4}{pixels}$ ($\SI{0.3}{\angstrom}$). The overlay in the top-right corner is an atomic model of a CuPcCl\textsubscript{16} molecule. (c) An ADF STEM simulation of a damaged CuPcCl\textsubscript{16} molecule after all chlorine has been lost due to electron irradiation. (d) ADF STEM simulation of an intact CuPcCl\textsubscript{16} molecule. The right half is overlayed with an atomic model of the molecule. (e) Line profiles taken from the images in panels (b), (c) and (d) that show the intensity of a single carbon atom compared to two carbon atoms on top of each other for the experimental (b) and the simulated (c) image. For comparison, also the intensity of a simulated chlorine atom is shown. All profiles where normalized in a way that a single carbon atom has an intensity of 1. The locations where the line profiles where taken are indicated by light green lines in the respective images.
	}
\end{figure}

Also at even higher doses the structure of the damaged, cross-linked molecules does not change significantly once it has stabilized (see figure~\ref{fig:closeup}). If the molecules lie close enough together so that their edges touch, they form a continuous thin amorphous carbon film. Larger distances, however, are bridged by 1D-carbon chains as can be seen in the top-right corner of figure~\ref{fig:closeup}b and \ref{fig:chlorine_loss}a.

 As already mentioned earlier we believe that the prevalent damage mechanism is the scission of the chlorine bonds. This is also supported by earlier studies which state that this bond is broken within $\SIrange{10}{100}{\femto\second}$ under electron irradiation~\cite{Fryer1984}. To further support this theory we studied the intensity of non-Cu atoms in a magnified image of the structure in more detail in figure~\ref{fig:chlorine_loss}. Panel c and d show a STEM simulation of a CuPcCl\textsubscript{16} on graphene, with the chlorine atoms removed and in its pristine state, respectively. Comparing the observed intensities to those in the experimental image (figure~\ref{fig:chlorine_loss}a) shows a good agreement of the experimentally observed intensities with the simulation without chlorine. Because of their high atomic number, chlorine atoms should be clearly visible in our ADF STEM image which can also be seen from the simulation in figure~\ref{fig:chlorine_loss}d. The overlay of an atomic model of CuPcCl\textsubscript{16} in the top-right corner of figure~\ref{fig:chlorine_loss}b shows where we would expect the chlorine atoms being located in an undamaged molecule. They are far enough away from the Cu core of the molecule so that they are not hidden by the blooming effect of this much brighter atoms.
 But also in the thin regions far away from the Cu atoms we can observe some bright atoms in the structure. To reveal their type, we took a line profile from one of the brightest non-Cu atoms in a thin region (figure~\ref{fig:chlorine_loss}b) and compared this profile to line profiles taken from the simulations in figure~\ref{fig:chlorine_loss}c and d. The result is shown in figure~\ref{fig:chlorine_loss}e. Clearly, the experimental profile matches the simulation without chlorine much better than the one with chlorine. A chlorine atom, under our imaging conditions, appears more than six times brighter than a carbon atom in graphene, whereas the brightest atoms in the experimental image and the simulation without chlorine are less than three times as bright as graphene. The brightness differences in the non-Cu atoms (which are most likely mostly carbon or nitrogen) occur due to the partially coherent imaging conditions given by the choice of detector angles in our experiment~\cite{Krivanek2012}.
 The STEM simulations were carried out with the package QSTEM~\cite{Koch2002} with zero probe aberrations. All other microscope parameters were set to values resembling our experimental conditions. The resulting images were blurred with a Gaussian kernel with a sigma of $\SI{0.4}{\angstrom}$ to account for the finite source size of the real microscope and other instabilities~\cite{Kirkland2010}.

\begin{figure}
	\centering
	\includegraphics[width=1.\textwidth]{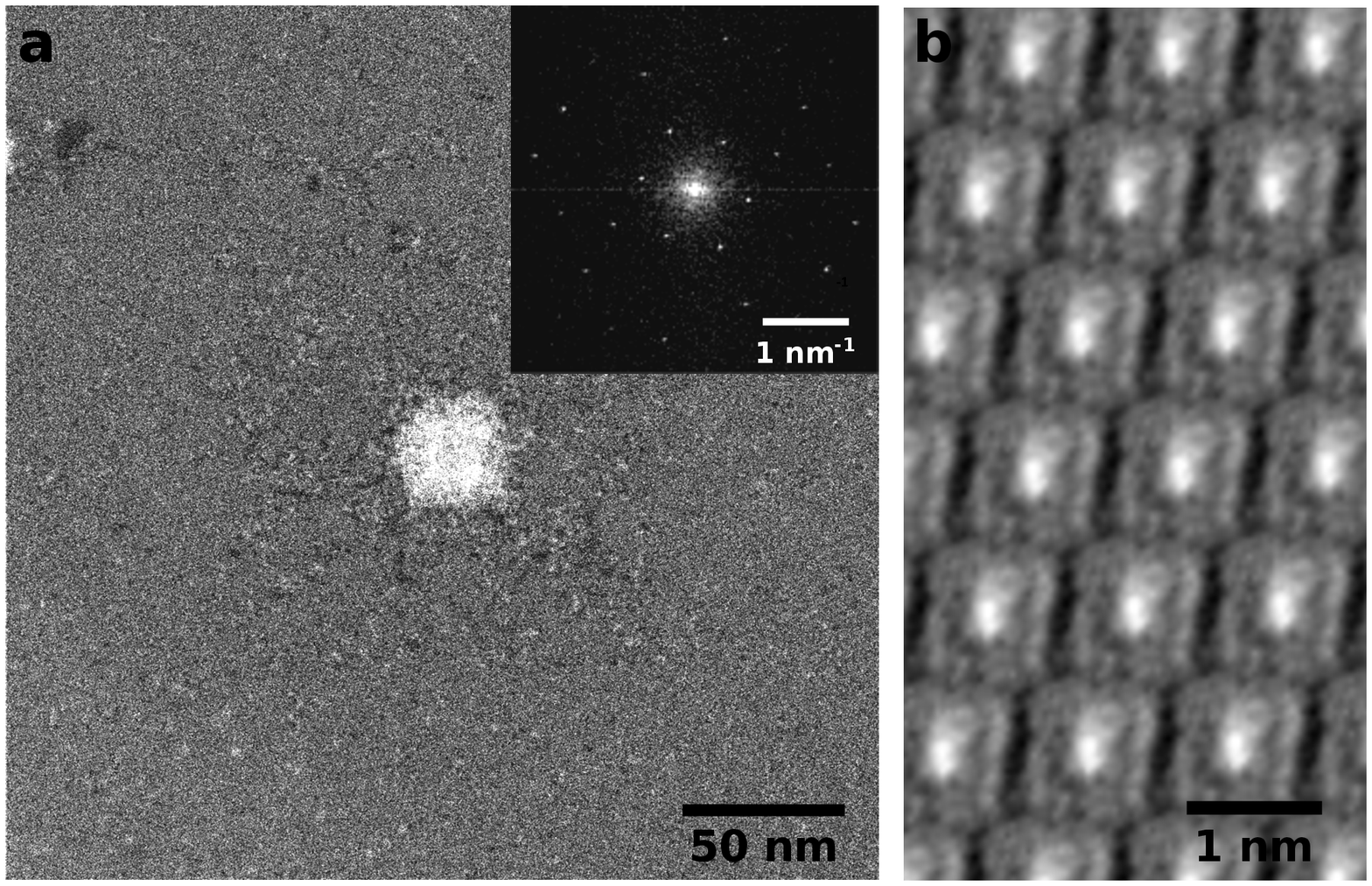}
	\caption{\label{fig:sampling_full}(a) Low-dose image of a monolayer CuPcCl\textsubscript{16} crystal. The bright square in the center is accumulated contamination and damaged molecules which was introduced by focusing in that spot. The electron dose is $\SI{\sim 160}{\electrons\per\angstrom\squared}$. (b) Translationally averaged and periodically repeated unit cell of the bottom-right quadrant of the image in a.}
\end{figure}

To image the undamaged CuPcCl\textsubscript{16} crystal the dose has to be reduced significantly. The two images in figure~\ref{fig:overview} indicate that the required dose should lie between $\SIrange[range-phrase=\text{ and }]{15}{500}{\electrons\per\angstrom\squared}$. To achieve this we jumped to the center of a previously not irradiated CuPcCl\textsubscript{16} crystal and focused there with a very small field of view. Then, a fast scan was acquired with a much larger field of view to capture the pristine state of the molecular crystal. The result is shown in figure~\ref{fig:sampling_full}a. The white square in the center consists of agglomerated destroyed molecules and hydrocarbon contamination introduced by focusing in this area. The rest of the image still contains large parts of the undamaged molecular crystal as shown by the power spectrum (inset of figure~\ref{fig:sampling_full}a). The power spectrum clearly shows the typical diffraction spots for this type of 2D-crystal lattice as it has been observed earlier employing STM~\cite{Scheffler2013}. 
To obtain a real-space representation of the molecules we used the diffraction spot positions to get a start value for the unit cell size and orientation of the crystal. Then the image was cut into those unit cells and all of them were averaged. In the next step the lattice parameters were optimized manually with visual feedback from the translationally averaged unit cell. Figure~\ref{fig:sampling_full}b shows the periodically repeated unit cell after manual optimization. This image was created only from the bottom-right quadrant of the raw image because the diffraction spots in the power spectrum of this quadrant were much more intense than in the other three quadrants. 

\begin{figure}
	\centering
	\includegraphics[width=1.\textwidth]{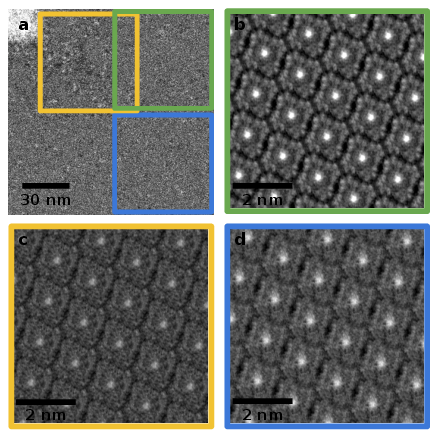}
	\caption{\label{fig:sampling_patches}(a) Bottom-right quadrant of the image in figure~\ref{fig:sampling_full}a. The colored squares show the sub-regions whose translational averaged unit cells are shown in panels b-c. (b) Sub-region with the best crystallinity. The molecular arrangement, the central copper atoms and even the 16 chlorine atoms around the edges of the molecules are clearly visible. (c) Translational average of a sub-region close to the focus spot (bright square). (d) Sub-region with worse crystallinity than the one in panel b. The copper atoms and the molecular arrangement are still visible but the chlorine atoms are not.}
\end{figure}

Since the crystallinity of the 2D molecular layer varies across the field of view,  we divided the bottom-right quadrant of the raw image into even smaller patches and applied the translational averaging method to each of them individually. As can be seen from figure~\ref{fig:sampling_patches}, there are significant differences in the results from different sub-areas: In panels c and d, only the central copper atoms and the molecular arrangement are visible. In contrast the shape of the individual molecules and even the 16 chlorine atoms around their edges are clearly visible in panel b. The crystal structure corresponds well with what has been reported earlier, both for monolayer crystals with STM and bulk crystals with (S)TEM~\cite{Haruta2012,Scheffler2013,Stock2015,Haruta2008,Yoshida2015,Uyeda1972}.
The critical electron dose, however, is much lower for our monolayer compared to what has been reported for bulk crystals~\cite{Kurata1992,Clark1980,Smith1986,Egerton2012,Fryer1984}. There are two main reasons for this: First, the critical electron dose was defined as the dose where the diffraction spot intensity has decayed to $\dfrac{1}{e}$. This definition is not sensitive to individual molecules being damaged but rather to the degradation of the long-range order in the crystal. The authors of Ref.~\cite{Clark1980} also state that damage to individual molecules occurs already at significantly lower doses, but due to subsequent crosslinking between adjacent molecules the long-range order is retained. Second, in bulk crystals the so-called "cage effect" plays an important role in the damage mechanism~\cite{Smith1986}: If a bond is broken, the created radical cannot escape the crystal (which is especially true for large atoms like chlorine) and therefore there is a high probability for recombination. This also explains why an EELS study on this material only showed a small decrease in chlorine content of the crystal even at doses as high as $\SI{\sim 20000}{\electrons\per\angstrom\squared}$~\cite{Kurata1992}. Thickness-dependent critical dose measurements also support the "caging effect" theory as the critical dose depends linearly on the crystal thickness~\cite{Fryer1984}. In our monolayer study this effect is completely absent, which explains why the dose that was needed for imaging the undamaged molecular crystal was only $\SI{\sim 160}{\electrons\per\angstrom\squared}$.

\section{Conclusions}
In summary, we present a study on the radiation damage process and critical dose in monolayer CuPcCl\textsubscript{16} crystals. In contrast to studies on bulk crystals, the damage mechanism and subsequent reactions that were proposed earlier can be directly imaged. We demonstrate that the main damage mechanism is a loss of the chlorine atoms which occurs instantly upon electron irradiation exceeding $\SI{\sim 160}{\electrons\per\angstrom\squared}$. Also crosslinking between damaged molecules could be observed which leads to a stabilization of the structure and makes it possible to image it at high electron doses. Finally, we demonstrate that it is possible to obtain an atomically resolved image of the monolayer crystal using low-dose acquisition and translational averaging.

\section*{Acknowledgements}
We acknowledge support from the European Research Council (ERC) Project No. 336453-PICOMAT.

\section*{Author contributions statement}
A.M. carried out experiments. A.M. and C.K. did the analysis. A.M. and J.C.M wrote the paper. J.C.M. supervised the work. All authors reviewed the manuscript.

\section*{Additional information}
\textbf{Competing interests:} The authors declare no competing interests.

\clearpage

\bibliography{library}

\end{document}